\documentclass[preprint]{aastex}

\slugcomment{}
\usepackage{bm}
\usepackage{graphicx}

\shorttitle{H.Kigure et al.}
\shortauthors{Faraday Rotation Measure Predicted from MHD Model II}

\begin{document}

\title{Distribution of Faraday Rotation Measure \\
in Jets from Active Galactic Nuclei \\
II. Prediction from our Sweeping Magnetic Twist Model \\
for the Wiggled Parts of AGN Jets and Tails}

\author{Hiromitsu Kigure\altaffilmark{1,2},Yutaka Uchida\altaffilmark{2,3},\\
Masanori Nakamura\altaffilmark{2,4}, Shigenobu Hirose\altaffilmark{2,5},
and Robert Cameron\altaffilmark{2,6}}
\altaffiltext{1}{Kwasan and Hida Observatories, Kyoto University,
Yamashina, Kyoto 607-8471, Japan;kigure@kwasan.kyoto-u.ac.jp}
\altaffiltext{2}{Department of Physics, Science University of Tokyo, 1-3
Kagurazaka, Shinjuku-ku, Tokyo 162-8601, Japan}
\altaffiltext{3}{Deceased, August 17, 2002.}
\altaffiltext{4}{Jet Propulsion Laboratory, California Institute of Technology, 4800 Oak Grove Drive, Pasadena, CA 91109, USA}
\altaffiltext{5}{Department of Physics and Astronomy, Johns Hopkins
University, Baltimore, MD 21218-2686, USA}
\altaffiltext{6}{Max-Planck-Institut f\"ur Aeronomie,
Max-Planck-Stra{\ss}e 2, D-37191 Katlenburg-Lindau, Germany}

\begin{abstract}
 Distributions of Faraday rotation measure (FRM) and the projected
 magnetic field derived by a 3-dimensional simulation of MHD jets are
 investigated based on our "sweeping magnetic twist model". FRM and
 Stokes parameters were calculated to be compared with 
 radio observations of large scale wiggled AGN jets on kpc scales.
 We propose that the FRM distribution can be used to discuss the 
 3-dimensional structure of magnetic field around jets and the validity
 of existing theoretical models, together with the 
 projected magnetic field derived from Stokes parameters. In the
 previous paper, we investigated the basic straight part of AGN jets by
 using the result of a 2-dimensional axisymmetric simulation.
 The derived FRM distribution has a general tendency to have a gradient
 across the jet axis, which is due to the toroidal component of the
 magnetic field generated by the rotation of the accretion disk.
 In this paper,
 we consider the wiggled structure of the AGN jets by using the result
 of a 3-dimensional simulation. Our numerical results
 show that the distributions of FRM and the projected magnetic field
 have a clear correlation with the large scale structure of the jet
 itself, namely, 3-dimensional helix. Distributions, seeing the jet from
 a certain direction, show a good matching with those in a part of 3C449 
 jet. This suggests that the jet has a helical structure and that the
 magnetic field (especially the toroidal component) plays an important
 role in the dynamics of the wiggle formation because it is due to a
 current-driven helical kink instability in our model.
\end{abstract}


\section{Introduction}
To explain the formation of active galactic nucleus (AGN) jets and
other astrophysical jets, various models have been proposed. Among them,
the magnetohydorodynamic (MHD) model is one of the most promising
models, since it can explain both the acceleration and the collimation
of the jets (see, e.g., Meier, Koide, \& Uchida 2001, and references
therein). Lovelace (1976) and Blandford (1976) first proposed the
theoretical model of the magnetically driven jet from accretion disks,
and Blandford \& Payne (1982) discussed magneto-centrifugally driven
outflow from a Keplerian disk in steady, axisymmetric and self-similar 
situation. Time-dependent, 2-dimensional axisymmetric simulations were
performed by Uchida \& Shibata (1985), Shibata \& Uchida (1986), Uchida
\& Shibata (1986). They pointed out that large amplitude torsional
Alfv\'en waves (TAWs) generated by the interaction between the
accretion disk and a large scale magnetic field play an important
role. The toroidal magnetic field propagates along the large scale
magnetic field while squeezing it into a collimated jet-shape by the
pinching effect of the Lorentz force.
In this paper, we refer this model as a "sweeping magnetic twist
model". After these papers, many authors have performed time-dependent,
2-dimensional axisymmetric simulations (e.g. Stone \& Norman 1994,
Ustyugova et al. 1995, Matsumoto et al. 1996, Ouyed \& Pudritz 1997,
Kudoh, Matsumoto, \& Shibata 1998).

Using the numerical data of MHD models,
observational quantities such as Faraday rotation measure (FRM) or 
Stokes parameters have been derived 
to be compared with observations of AGN jets: 
Laing (1981) computed the
total intensity, the linear polarization, and the projected magnetic
field distributions, assuming some simple magnetic field configurations 
and high energy particle distributions in the cylindrical jet. Clarke, 
Norman, \& Burns (1989) performed 2-dimensional MHD
simulations in which a supersonic jet with a dynamically passive helical
magnetic field was computed, and derived distributions of the total 
intensity, the projected electric field, and the linear polarization. 
Hardee \& Rosen (1999) calculated the total intensity and the projected
magnetic field distributions, using 3-dimensional MHD simulations of
strongly magnetized conical jets. Hardee \& Rosen (2002)
calculated the FRM distribution and discussed
that the observation of the radio source 3C465 in Abell cluster A2634
(Eilek \& Owen 2002) suggests helical twisting of the flow.

FRM is given by the integral of $n_e B_\parallel$ along 
the line-of-sight between the emitter and the observer (where
$B_\parallel$ is the line-of-sight component of the magnetic field, and
$n_e$ is the electron density there). It is, in principle, not possible
to specify which part on the line-of-sight the contribution comes
from. However, in recent high-resolution radio observations (e.g. Eilek
\& Owen 2002, Asada et al. 2002), the FRM distribution seems to have
good correlation with the configuration of the jet;
this suggests that the FRM
variation is due to the magnetized thermal plasma surrounding
the emitting part of the jet. In fact, sharp FRM gradients seen in
3C273 can not be produced by a foreground Faraday screen
(Taylor 1998, Asada et al. 2002).
If this is the case, 
we can get a new information, that is, the line-of-sight component of 
the magnetic field, and thus can predict 
the 3-dimensional configuration of the
magnetic field around the jet, together with the projected magnetic field.

Uchida et al. (2004) (hereafter paper I) carried out a 2-dimensional
axisymmetric simulation, and investigated the model counterpart
distributions of FRM and the projected magnetic field in the basic
straight part of AGN jets. 
It was described how a systematically helical field configuration is
produced in the "sweeping magnetic twist model". It was also shown
as a result that the model can reproduce fairly well the
characteristic distribution of FRM having the gradient across the
jet axis (Perley, Bridle, \& Willis 1984, Asada et al. 2002, Gabuzda \&
Murray 2003). The systematic FRM gradient was caused by the gradient of
the line-of-sight component of the magnetic field. This means that the
existence of the helical magnetic field (and the propagation of
TAWs) is plausible. On the basis of this success, in this paper the
treatment is advanced to the non-axisymmetric situation, the wiggled
structure, in AGN jets.

The morphological structures in AGN jets, such as "wiggles (kinks)" or
"bends" are frequently seen not only on kpc scales, but also on pc
scales and smaller (Hummel et al. 1992). Such a helical distortion might
be caused either by plasma instabilities or precession of jet ejection
axis due to the gravitational interaction between binary Black Holes
(BBHs) (Begelman, Blandford, \& Rees 1980), BH/disk, or
galaxies. Magnetically driven jets possess a toroidal field component,
which is equivalent to an axial electric current, and such
"current-carrying" jets are susceptible to MHD instabilities, and
moreover, Kelvin-Helmholtz instabilities must be also taken into
account for jet dynamics. MHD instabilities are usually divided into
pressure-driven instabilities and current-driven instabilities (Bateman
1980). Kelvin-Helmholtz instabilities have been considered for past
decades by many theoretical and numerical work (for reviews, see
Birkinshaw 1991; Ferrari 1998, and references therein).
One of the promising possibilities is a magnetic mechanism based on the
"sweeping magnetic twist model". It was shown by 3-dimensional MHD
simulations that the formation of the wiggled structure can be explained
by the current-driven helical kink instability (Todo et al. 1993). In
these simulations, the magnetic field was a force-free helical field
from the beginning and the propagation of TAWs was not dealt with.
Nakamura, Uchida, \& Hirose (2001) extended the treatment of the
"sweeping magnetic twist model" to the part far from the gravitator and
the accretion disk. They investigated the behavior of TAWs propagating
far from the AGN core. They showed that the current-driven helical
kink instability can explain the production of the observed wiggles.
If the mechanism of the wiggle formation is clarified, it can become a
clue to the physics of the jet formation.

In this paper, we calculate FRM, the projected magnetic field, and 
the total intensity from the numerical data of a MHD simulation 
based on our ``sweeping magnetic twist model'', and discuss these model 
counterparts comparing with an observation. Here we consider the 
wiggled structure of the jet, and thus use the same kind of
3-dimensional simulation as in Nakamura et al. (2001).
In section 2, we describe the application of the model to the distant
part of the jet from the AGN core, and show the formation of a
helical structure of the jet by the current-driven helical kink
instability. We introduce the method to calculate model counterparts of
observational quantities in section 3, and show the results in section
4. We discuss comparison between the calculated distributions and
observed ones in section 5. The conclusion is summarized in section 6.

\section{Numerical Simulation} 
The ``sweeping magnetic twist model'' was proposed for star-forming jets
(Uchida \& Shibata 1985, Shibata \& Uchida 1986), and it has since then
been extended to the case of AGN jets (Uchida \& Shibata
1986, Matsumoto et al. 1996, Uchida et al. 1999, Uchida et al. 2000,
Nakamura et al. 2001) as described in paper I in some detail.
In paper I, we confined ourselves to the 
straight part of the jet as a first step, by using the 2-dimensional 
axisymmetric model. 
In the present paper, we extend our treatment to the wiggled part of our
3-dimensional jet model, in order to compare the model counterpart with
the observation. In this section, we describe the formation of the
wiggled structure due to the current-driven helical kink instability.
\subsection{Brief Explanation of the Model --- Assumptions and Basic
Equations}
\begin{figure}
\figurenum{1}
\epsscale{1.0}
\plotone{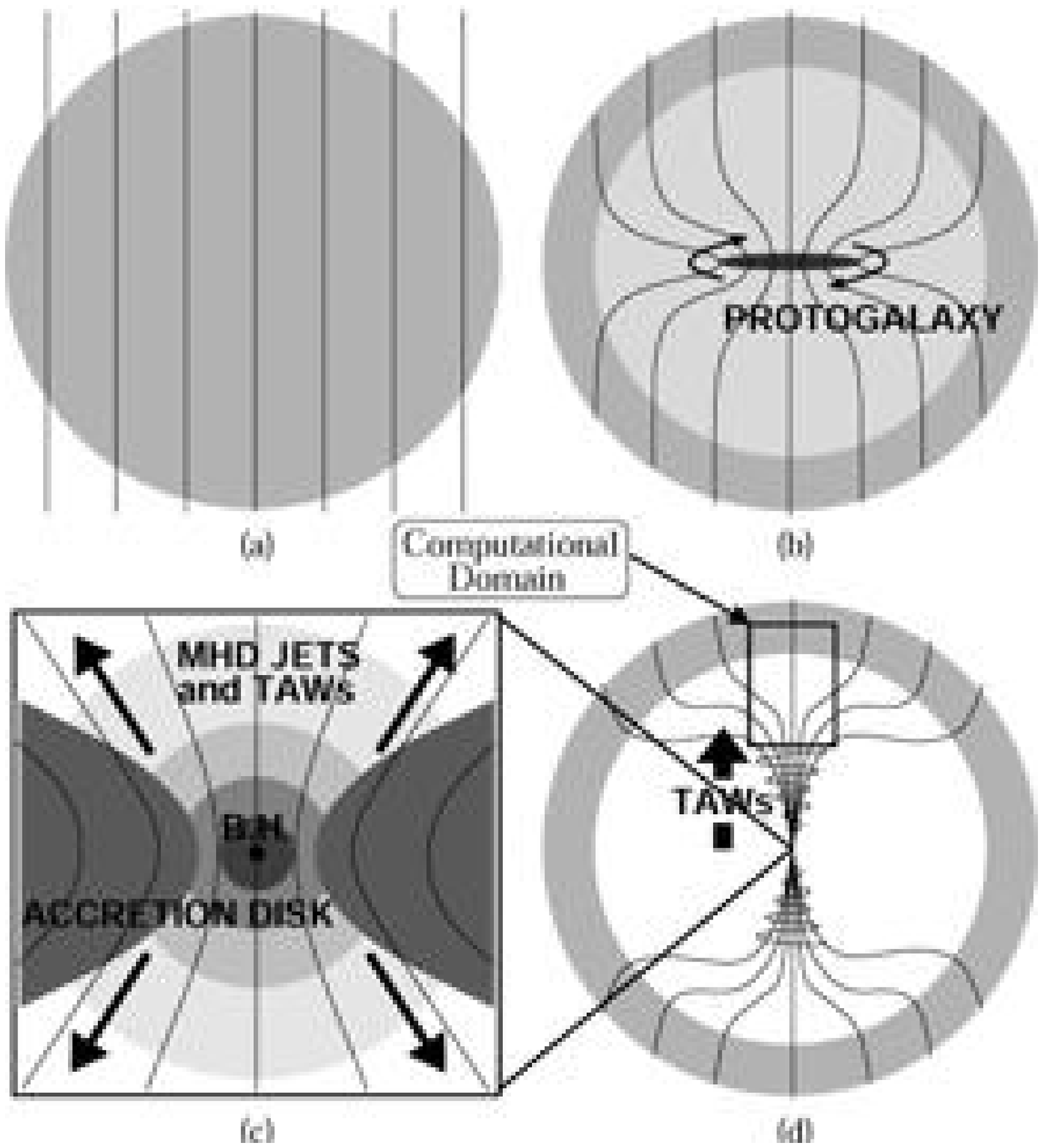}
\caption{Schematic picture of our model. (a) From the intergalactic
 medium with a large scale magnetic field, (b) the AGN core region and
 the disk are formed due to the gravitational contraction of mass with
 the magnetic flux conserved. (c) Torsional Alfv\'en waves generated by
 the interaction between the accretion disk and the large scale magnetic
 field (d) propagate along the magnetic field. The computational domain
 which we assumed is shown.
\label{FIG01}} 
\end{figure}

Figure \ref{FIG01} is the schematic picture of a physical situation we
treat in this paper. We consider a primordial large scale magnetic
field, and that it is squeezed under an assumption of frozen-in magnetic
flux due to the gravitational contraction to form the central AGN core
(Figure \ref{FIG01}a, b). During the contracting process of the
magnetized gas, the toroidal component of the magnetic field is
continuously created by the rotation of the intergalactic medium having
the angular momentum. As a result, large amplitude non-linear torsional
Alfv\'en waves (TAWs) begin to propagate out along the large scale
poloidal field. We expect that this process reaches its maximum as the
accretion disk is formed, and the interaction of the rotating accretion
disk and the poloidal magnetic field penetrating it reaches its maximum
phase (Figure \ref{FIG01}c). And also, the MHD jets powered by TAWs are
emitted in opposite directions along the rotation axis of the disk which
is the direction of the original large scale poloidal magnetic field.

In the present paper, we concentrate our attention on the distant part
of the jet from the AGN core. The TAWs finally encounter the ambient
medium with a higher density surrounding the large spherical "cavity"
from which the mass fell toward the center in the gravitational
contraction (Figure \ref{FIG01}d). The HST optical and MERLIN radio
observations of a few hundreds pc-scale 3C264 jet indicate that the jet
has a wiggled structure as a result of interacting with a circular
optical cold (dense) "ring", which may be a true ring, a shell, or a
filled spheroid (Baum et al. 1997). This observational result indicates
the interaction between the jet and the ambient medium will affect the
jet dynamics and morphology. The existence of denser material ahead of
the propagating TAWs causes the deceleration of MHD jets due to the
decreasing local Alfv\'en velocity; the accumulation of the toroidal
component of the magnetic field occurs, which may lead to the
current-driven helical kink instabilities through this non-linear process
(Nakamura et al. 2001). Even in the case of decreasing density
atmospheres, the MHD jets would be subject to the current-driven kink
distortions (M. Nakamura \& D. L. Meier, in preparation).

The computational domain we report here is also shown in Figure
\ref{FIG01}(d), and it includes a part of "squeezed" poloidal field
near the edge of the density "cavity". We assume the poloidally
magnetized intergalactic medium with a varying Alfv\'en velocity
distribution ({\it i.e.}, gradually decreasing as the TAWs propagate)
due to the existence of a denser shell which represents the edge of the
"cavity" . A continuous cylindrical MHD inflow, powered by TAWs into the
"evolved" region of the domain, is specified for all time in the lower
"boundary zone". We note that a quasi-stationary Poynting-flux-dominated
flow is injected into our calculated region throughout the time
evolution.

We adopt the ideal MHD as the governing equations.
\begin{equation}
{\frac{\partial \rho}{\partial t}}+\nabla \cdot (\rho \bm{v})=0
\end{equation}
\begin{equation}
\rho \frac{\partial \bm {v}}{\partial t} + \rho (\bm{v} \cdot \nabla) \bm{v} 
+ \nabla p - \frac{1}{4\pi} (\nabla \times \bm{B}) \times \bm{B} =0
\end{equation}
\begin{equation}
\frac{\partial p}{\partial t} + \nabla \cdot (p \bm{v}) - (1 - \gamma) p 
\nabla \cdot \bm{v} = 0
\end{equation}
\begin{equation}
{\frac{\partial \bm{B}}{\partial t}} - \nabla \times 
(\bm{v} \times \bm{B}) = 0
\end{equation}
where $\rho, p, \bm{v}$ are the density, pressure, and velocity of the
gas, and $\bm{B}$ is the magnetic field. $\gamma$ represents the ratio
of specific heats and is equal to 5/3 in this paper.
In paper I, we introduced the effect of gravity in eq. (2) in order to
consider the launching and accelerating of MHD jets near
the AGN core. We neglect the gravity term in this equation, and
concentrate on the dynamics of the propagating MHD jets in this paper.
We normalize all physical quantities by the typical Length $L_0$,
typical density $\rho_0$, and typical velocity $V_0$. $L_0$ is ten times
as long as the radius with the maximum velocity of the circular
motion. $\rho_0$ and $V_0$ are the density and the Alfv\'en velocity at
the origin respectively.
The total computational domain is taken to be 
$x_{\rm min} \le x \le x_{\rm max}$, 
$y_{\rm min} \le y \le y_{\rm max}$, and 
$z_{\rm min} \le z \le z_{\rm max}$, where
$x_{\rm min}=y_{\rm min}=-5.10$, $x_{\rm max}=y_{\rm max}=5.10$, 
$z_{\rm min}=-1.00$, and $z_{\rm max}=12.09$.
The numbers of grid points in the simulation reported here are
($N_{x} \times N_{y} \times N_{z}$) $=$ $(243 \times 243 \times 655)$, 
where the grid points are distributed non-uniformly in the $x$, $y$, and
$z$ directions. The grid spacing is uniform, ($\Delta x, \Delta y,
\Delta z$) $=$ ($0.015, 0.015, 0.015$) for $|x| \leq 0.99$, $|y| \leq
0.99$, and $z \leq 8.0$, and then stretched by 5 \% per each grid step
for the regions $|x|>0.99$, $|y|>0.99$, and $z>8.0$.

\begin{figure}
\figurenum{2}
\epsscale{0.8}
\plotone{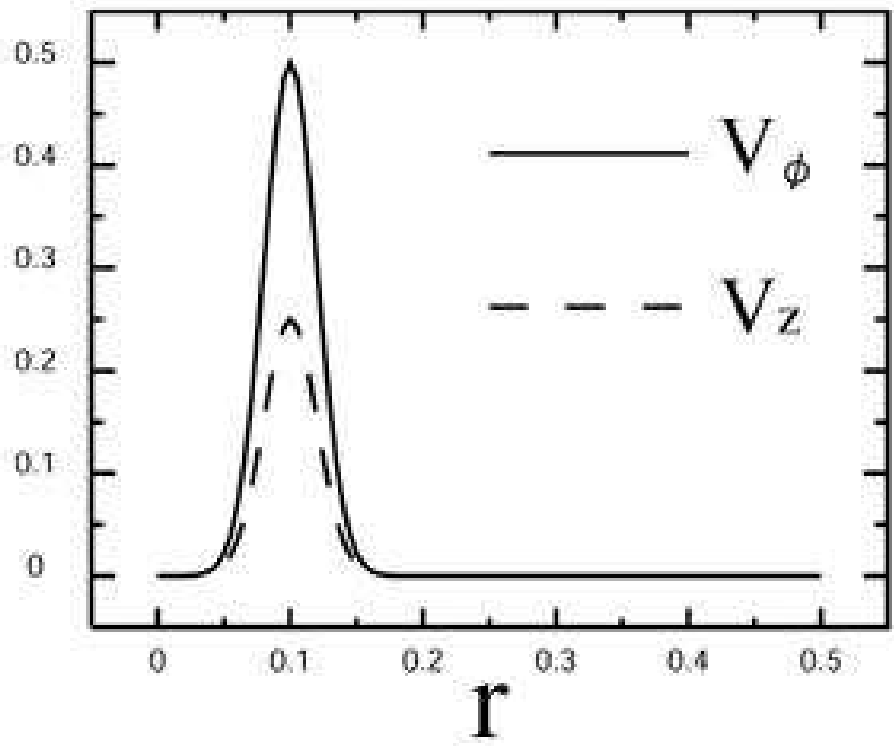}
\caption{The radial profile of imposed circular motion (solid line) and
 inflow (dotted line) at the lower boundary
\label{FIG02}} 
\end{figure}

\begin{figure}
\figurenum{3}
\epsscale{0.8}
\plotone{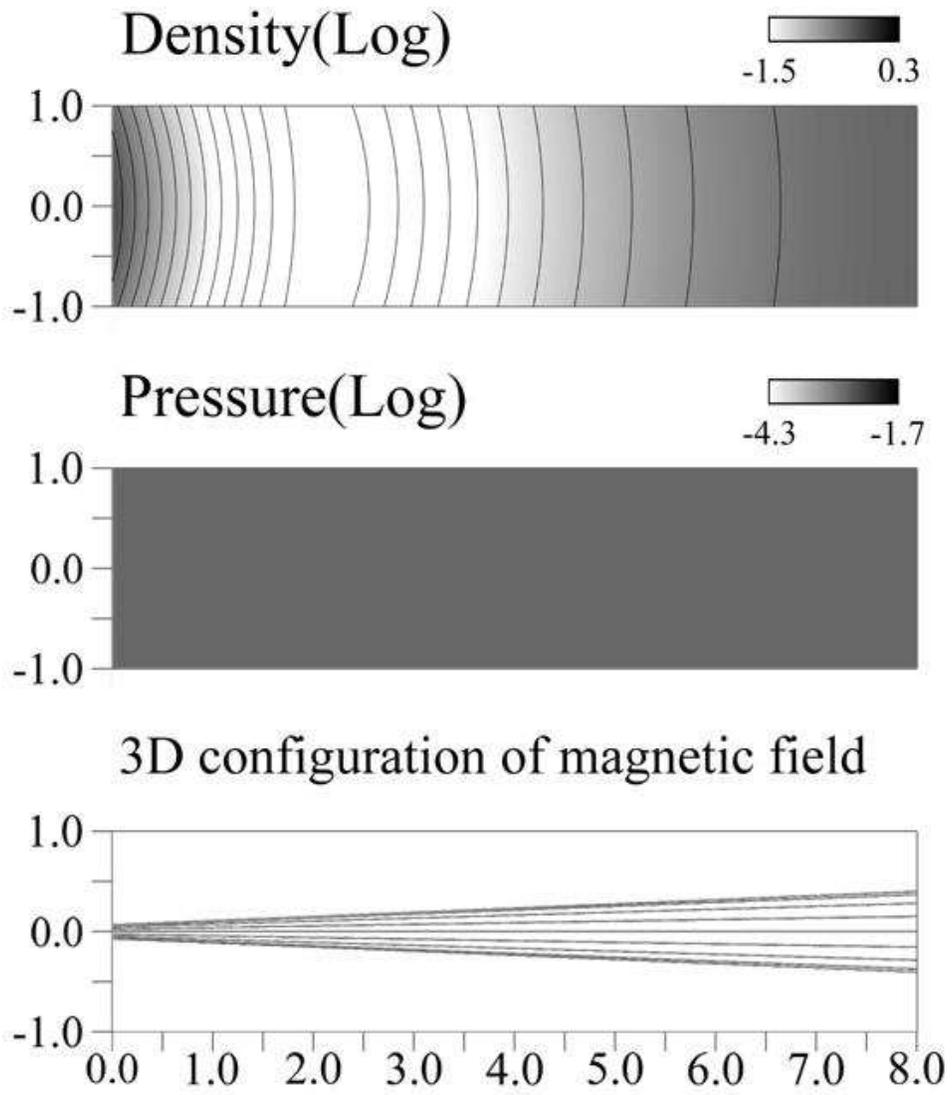}
\caption{The initial distributions of (a) the density,
 (b) the pressure (initially constant) on the $y-z$ plane ($-1.0 \leq y 
 \leq 1.0$, $0.0 \leq z \leq 8.0$) at
 $x=0.0$, and (c) the projected 3-dimensional configuration of the
 initial magnetic field.
\label{FIG03}} 
\end{figure}

Initial conditions are almost the same as those of Nakamura et
al. (2001), but the radial profiles of $V_\phi$ and $V_z$ (the circular
motion and inflow imposed at the lower "boundary zone") in this paper are
different from those in Nakamura et al. (2001). The profiles in this
paper are shown in Figure \ref{FIG02}. 
The initial distributions of the density and the pressure of the 
gas, and the 3-dimensional configuration of selected magnetic lines
of force, are shown in Figure \ref{FIG03}.
This figure displays the distributions of the range of $-1.0 \leq y \leq 
1.0$, $0.0 \leq z \leq 8.0$.
$B_{z}$ decreases gradually along the $z$-axis and tend to $\sim z^{-2}$
for the upper region, and $\rho$ also decreases in the same way, but
increases again toward the upper boundary. We assume a constant pressure
throughout the computational domain and the plasma $\beta$ (the ratio of 
gas pressure to magnetic pressure) at the origin is $0.01$. 
The value of the plasma $\beta$ increases along the $z$-axis and exceeds
unity at $z=3.8$.
In the decreasing region ($\rho \ {\rm and} \ \mbox{\bm$B$}$) in the lower 
part, we have the constant Alfv\'en velocity ($V_{\rm A} \equiv 
|\mbox{\bm$B$}|/\sqrt{\rho}$), and gradually decreasing distribution of 
$V_{\rm A}$ as $\rho$ increases.
The formation of the wiggled structure is universal independent of the
imposed velocity profiles. The clearest difference in the results
between Nakamura et al. (2001) and this paper is the time which it takes
before the wiggled structure is formed.

\subsection{Time evolution}
\begin{figure}
\figurenum{4}
\epsscale{0.72}
\plotone{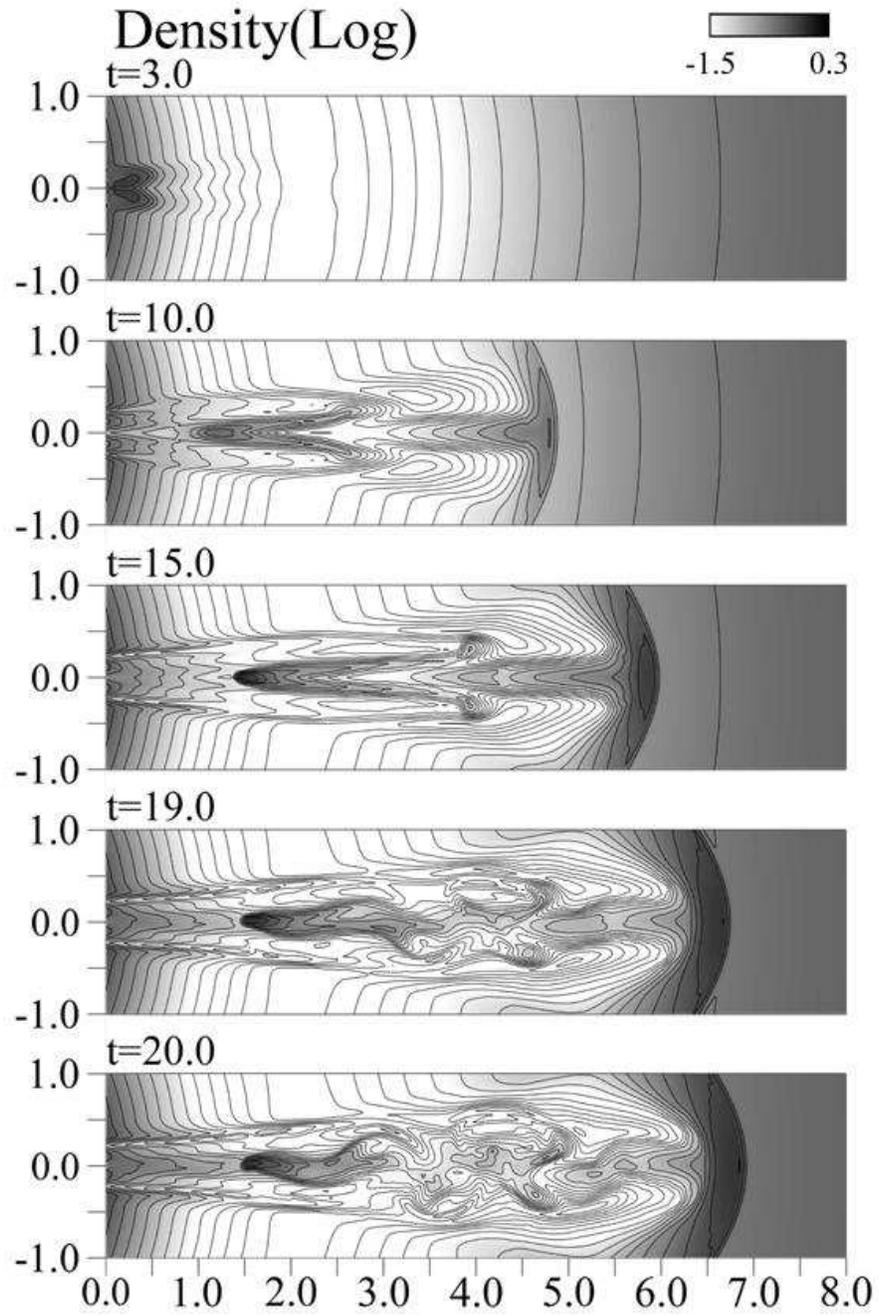}
\caption{Time evolution of the density on the $y-z$ plane at $x=0.0$
($-1.0 \leq y \leq 1.0$, $0.0 \leq z \leq 8.0$).
\label{FIG04}} 
\end{figure}

\begin{figure}
\figurenum{5}
\epsscale{0.72}
\plotone{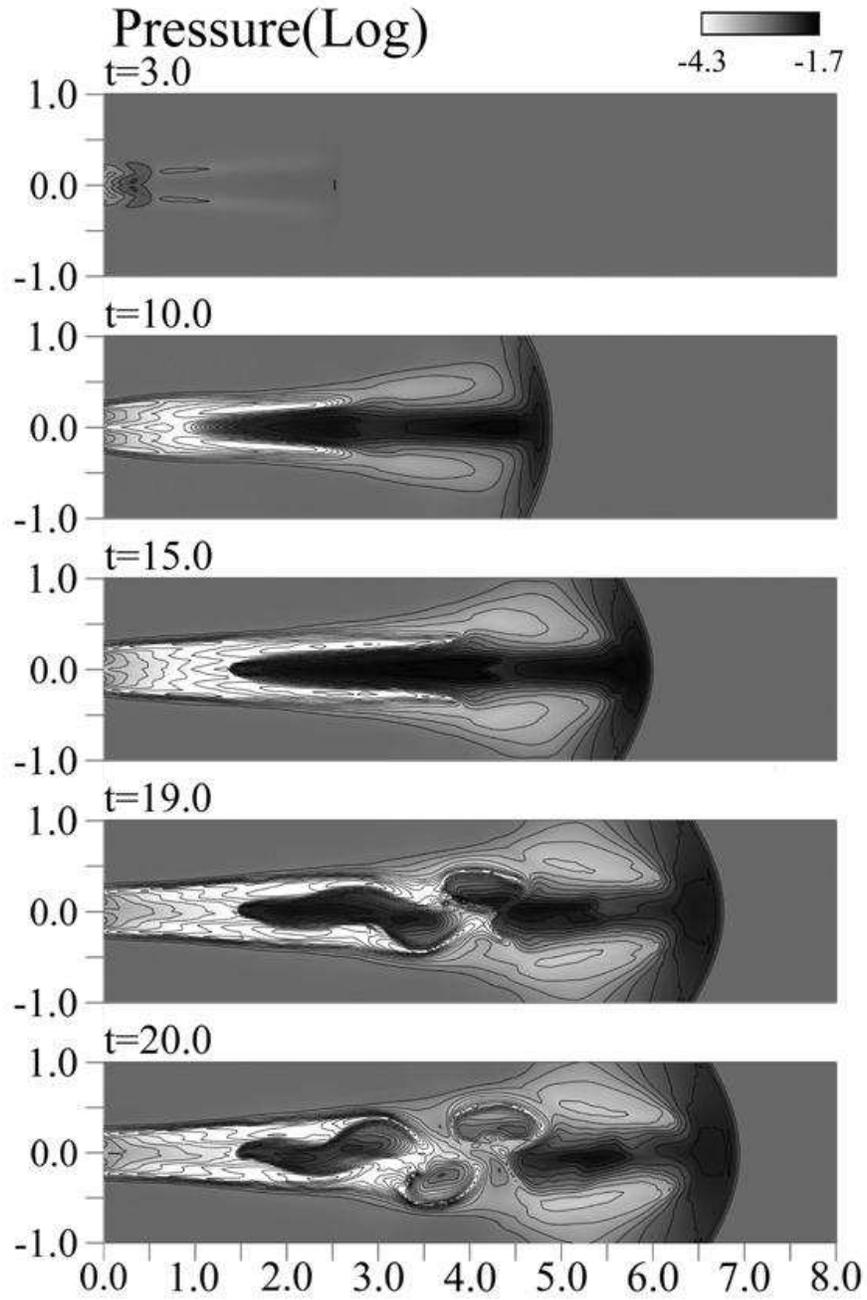}
\caption{Time evolution of the pressure on the $y-z$ plane at $x=0.0$.
\label{FIG05}} 
\end{figure}

\begin{figure}
\figurenum{6}
\epsscale{0.72}
\plotone{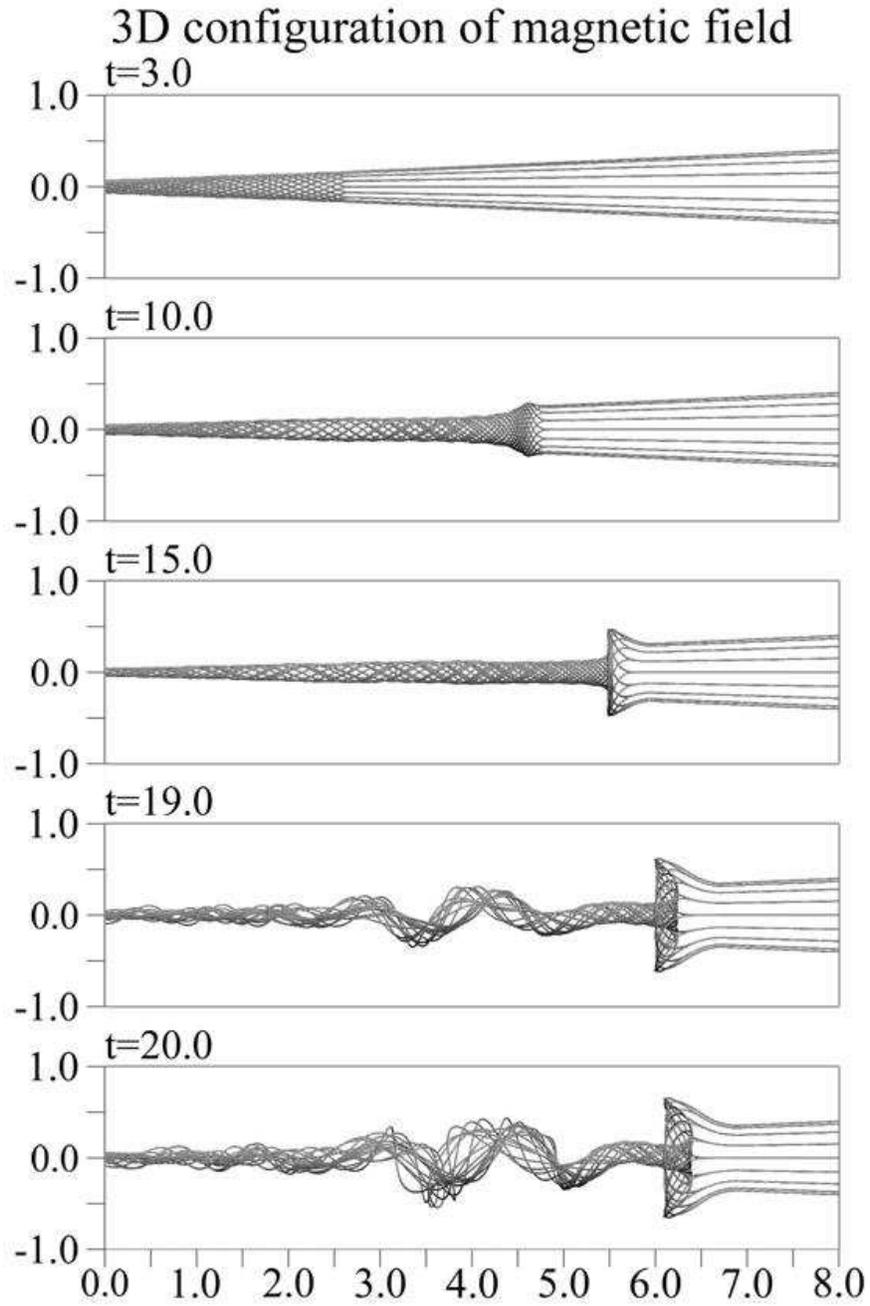}
\caption{Time evolution of the projected 3-dimensional configuration of
 selected magnetic lines of force.
\label{FIG06}}
\end{figure}

\begin{figure}
\figurenum{7}
\epsscale{1.0}
\plotone{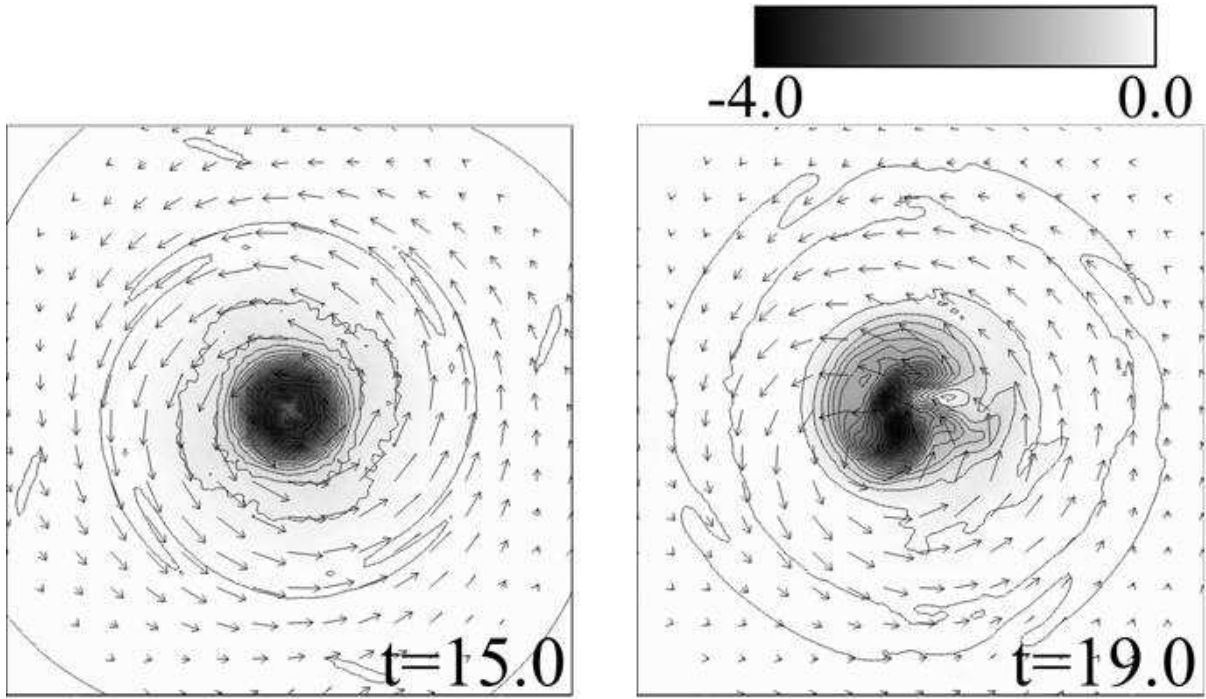}
\caption{Distribution in the $x-y$ plane at $z=4.25$. A grayscale and
 contour show the $z$-component of the current density and arrows show the
 velocity field.
\label{FIG07}}
\end{figure}

\begin{figure}
\figurenum{8}
\epsscale{0.25}
\begin{center}
\rotatebox{-90}{\plotone{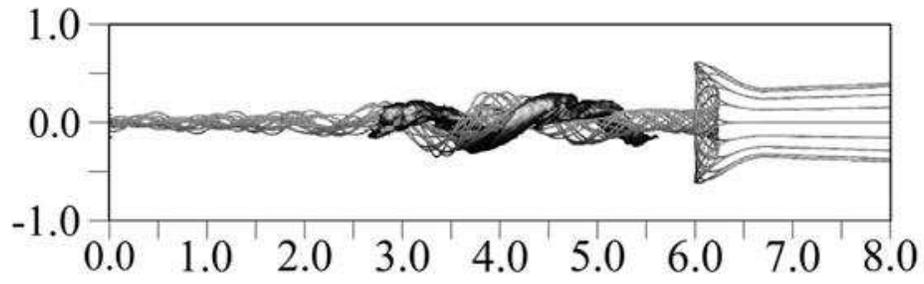}}
\caption{3-dimensional distributions of a typical iso-value surface
 of the specific power: $-J_z B_\phi V_r / \rho + J_\phi B_z V_r / \rho$
 and the selected magnetic lines of force at $t = 19.0$.
\label {FIG08}}
\end{center}
\end{figure}

We discuss the dynamical behavior of a typical case in our performed
3-dimensional simulations: the growth of a current-driven instability in
a MHD jet. Figure \ref{FIG04}, \ref{FIG05}, and \ref{FIG06} show the
time evolution of the density, pressure, and 3-dimensional configuration
of the selected magnetic lines of force in the $y-z$ plane ($-1.0 \leq y 
\leq 1.0$, $0.0 \leq z \leq 8.0$) at $x=0.0$. In the early stage ($t=3.0$), 
a strongly magnetized helical jet powered by TAWs advances with a constant 
velocity into the decreasing poloidally magnetized medium. Our MHD jets 
have a fast-mode MHD wave front, and a slow-mode MHD wave front closely 
beyond a contact surface. The front of the TAW (a fast-mode MHD wave) 
begins to be decelerated due to the gradually decreasing $V_{\rm A}$; the
accumulation of $B_{\phi}$ occurs behind this wave front ($t=10.0$). The
toroidal magnetic pressure gradient force $d (B_{\phi}^{2}/2) /dz$
becomes large and works effectively at this front, strongly compressing
the external medium along this propagating TAW front. The initially
"hourglass-shaped" magnetic field becomes "tucked up" behind this wave
front ($t=15.0$). At later stages the front becomes super-fast
magnetosonic and a strong bow shock (fast-mode MHD shock) is formed. The
accumulation of $B_{\phi}$ causes a concentration of the axial current
density $J_{z}$ near the central ($z$) axis (Figure \ref{FIG07} {\it
left}), and the magnetic pitch ($|B_{\phi}/rB_{z}|$) becomes large
inside the jet. The Lorentz force breaks the quasi-equilibrium balance
in the radial direction, and the jet is disrupted ($t=19.0 - 20.0$). We
have an asymmetric distribution of the current density and projected (in
a $x-y$ plane) velocity field seen in Figure \ref{FIG07} {\it
right}. There is a strong correlation between the 3-dimensional
configuration of the magnetic field and the distribution of specific
power due to the radial ($r$) component of the Lorentz force: $-J_{z}
B_{\phi} V_{r} /\rho + J_{\phi} B_{z} V_{r}/\rho$ (Figure
\ref{FIG08}). Therefore it is confirmed that the distortion of our MHD
jet is driven by the current-driven helical kink instability ($m=1$)
(Nakamura et al. 2001).

\section{Method of Calculation of Model Counterparts}
Using the numerical data of the 3-dimensional simulation explained
in the previous section, we computed the distributions of FRM, the
projected magnetic field, and the total intensity seen from different
viewing angles. 

We computed the FRM distribution by integrating $B_\parallel$
along the line-of-sight although in paper I we integrated
$n_e B_\parallel$. In this paper, we intend to consider the effect of
the deformed magnetic field configuration and to omit the effect of the
dense shell from consideration. To calculate Stokes parameters, we
assume the following:
(1)radiation processes are dominated by synchrotron radiation,
(2)synchrotron self absorption is negligible, (3)the spectral index,
$\alpha$, is equal to unity, and (4)the projected magnetic field is
perpendicular to the projected electric field.
The emissivity of the synchrotron radiation is given by 
$\epsilon = p |B \sin \psi|^{\alpha + 1}$,
where $B$ is the local magnetic field strength, $\psi$ is the
angle between the local magnetic field and the line-of-sight, and $p$ is
the gas pressure. In our simulation the relativistic particles are not
explicitly tracked, therefore we assume that the energy and number
densities of the relativistic particles are proportional to the energy
and number densities of the thermal fluid (Clarke et al. 1989, Hardee \&
Rosen 1999, 2002). The total intensity is then given by the integration
of the emissivity along the line-of-sight as 
$I = \int \epsilon ds $. Other Stokes parameters 
are given by $Q = \int \epsilon \cos 2 \chi^{'} ds$ and 
$U = \int \epsilon \sin 2 \chi^{'} ds$,
where the local polarization angle 
$\chi^{'}$ is determined by the direction of the local magnetic field 
and the direction of the line-of-sight. Using these $U$ and $Q$, the
polarization angle $\chi$ is given by 
$\chi = (1/2)\tan^{-1}(U/Q)$. Finally the projected magnetic field is 
determined from the polarization angle
$\chi$ and the polarization intensity $\sqrt{Q^2 + U^2}$.

Here we separate the Faraday rotation screen and the emitting region, 
and we performed the integrations only in the 
emitting region for Stokes parameters, 
and only in the Faraday rotation screen for FRM. 
We assumed this separation on the basis of the fact that 
linear dependence of the observed 
polarization angle on wavelength-squared holds in 
some observations (Perley et al. 1984, Feretti et al. 1999, 
Asada et al. 2002, Gabuzda \& Murray 2003); this would not be the case
if the Faraday rotation is caused in the emitting region (Burn 1966).
The emitting region and the Faraday rotation screen are deformed into
the 3-dimensional shape in the helical kink instability. Therefore we
trace the flux tube in its time development by the initial footpoints of
magnetic lines of force at the upper boundary of the simulation
region. This is reasonable because the footpoints don't move until a
wave front arrives. The region which is defined as the Faraday rotation
screen is almost corresponding to the magnetic tube to which the
circular motion is imposed.

\section{Results of the Numerical Observation}
\subsection{Before the formation of the wiggled structure}
\begin{figure}
\figurenum{9}
\epsscale{1.0}
\plotone{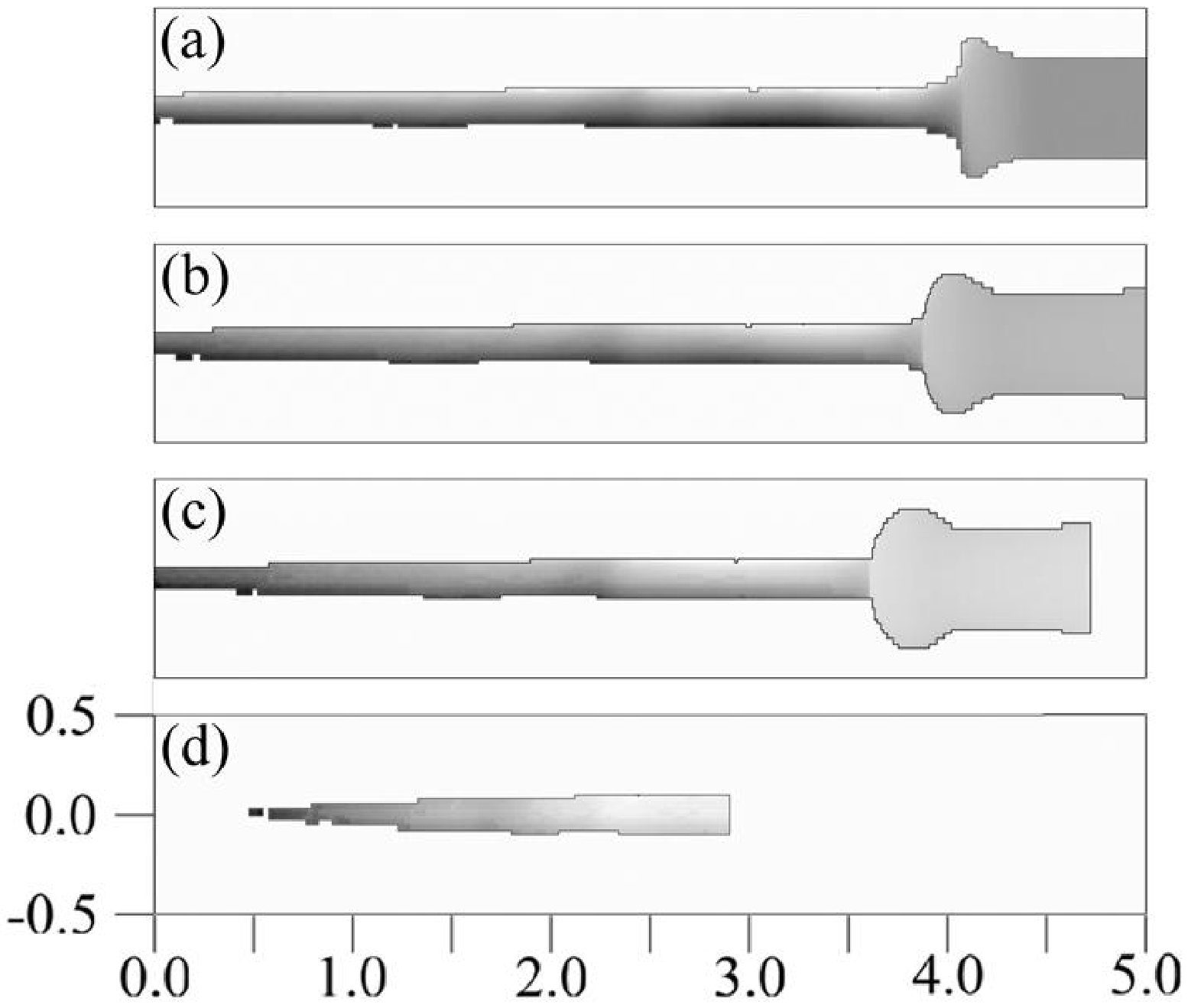}
\caption{Calculated model counterparts for the Faraday rotation measure
 distribution, when seen at (a) $90^\circ$, (b) $70^\circ$, (c)
 $55^\circ$, and (d) $30^\circ$, from the axis (ahead of the jet) at the
 time before the formation of the wiggled structure ($t = 15.0$).
 The data of the numerical observation exists inside the area surrounded
 by the black line.
\label{FIG09}}
\end{figure}

Figure \ref{FIG09} shows the FRM distributions at the time before the
formation of the wiggled structure ($t = 15.0$). Black in the map
corresponds to a maximum and white to a minimum in each viewing angle
case. The FRM distributions have the gradient in the direction
perpendicular to the jet axis independent of the viewing angle. This
gradient is caused by the toroidal magnetic field propagating as TAWs.

\begin{figure}
\figurenum{10}
\epsscale{1.0}
\plotone{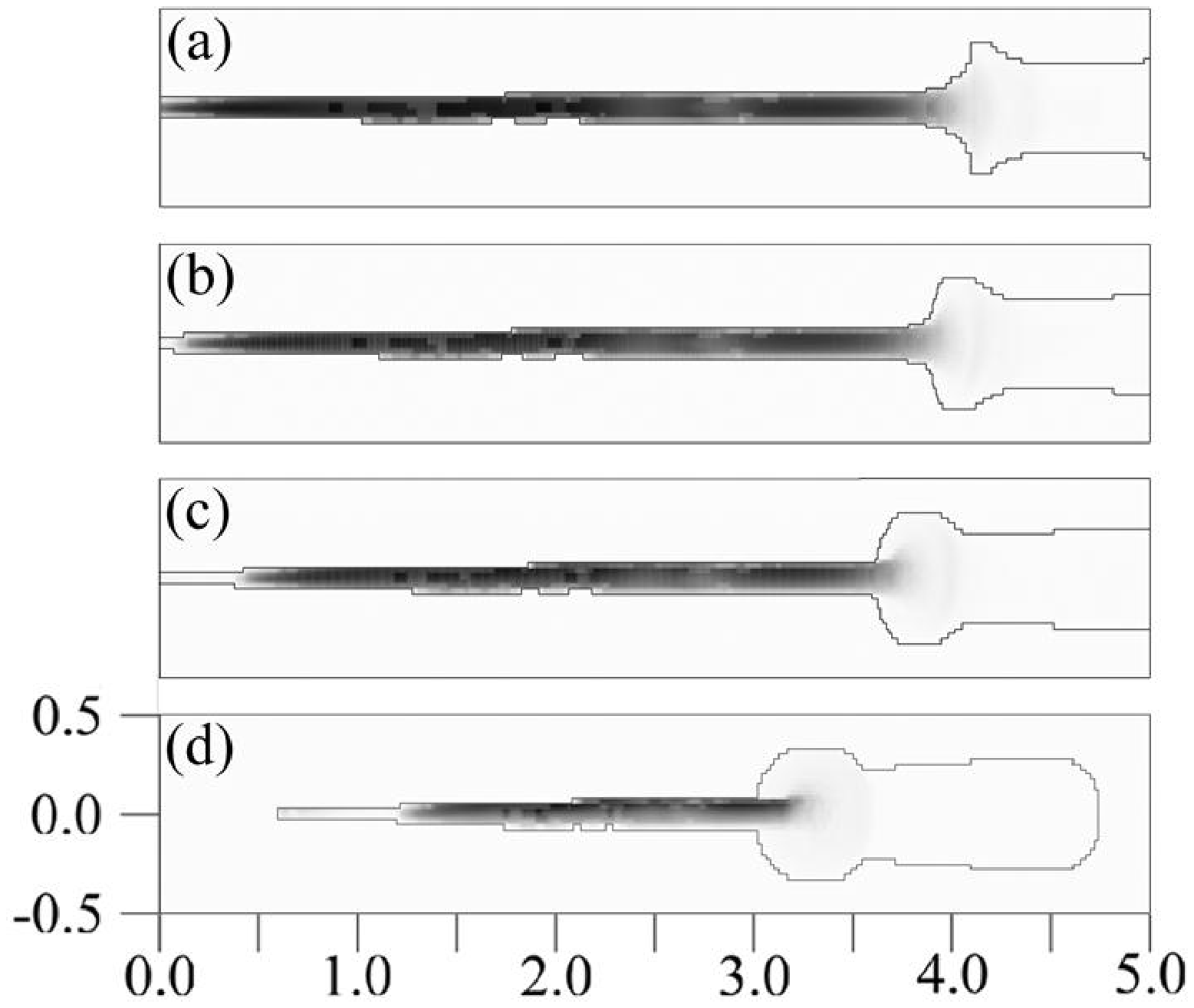}
\caption{Calculated model counterparts for the total intensity, when
 seen at (a) $90^\circ$, (b) $70^\circ$, (c) $55^\circ$, and (d)
 $30^\circ$, from the axis at $t = 15.0$.
 The data of the numerical observation exists inside the area surrounded
 by the black line.
\label{FIG10}}
\end{figure}

\begin{figure}
\figurenum{11}
\epsscale{1.0}
\plotone{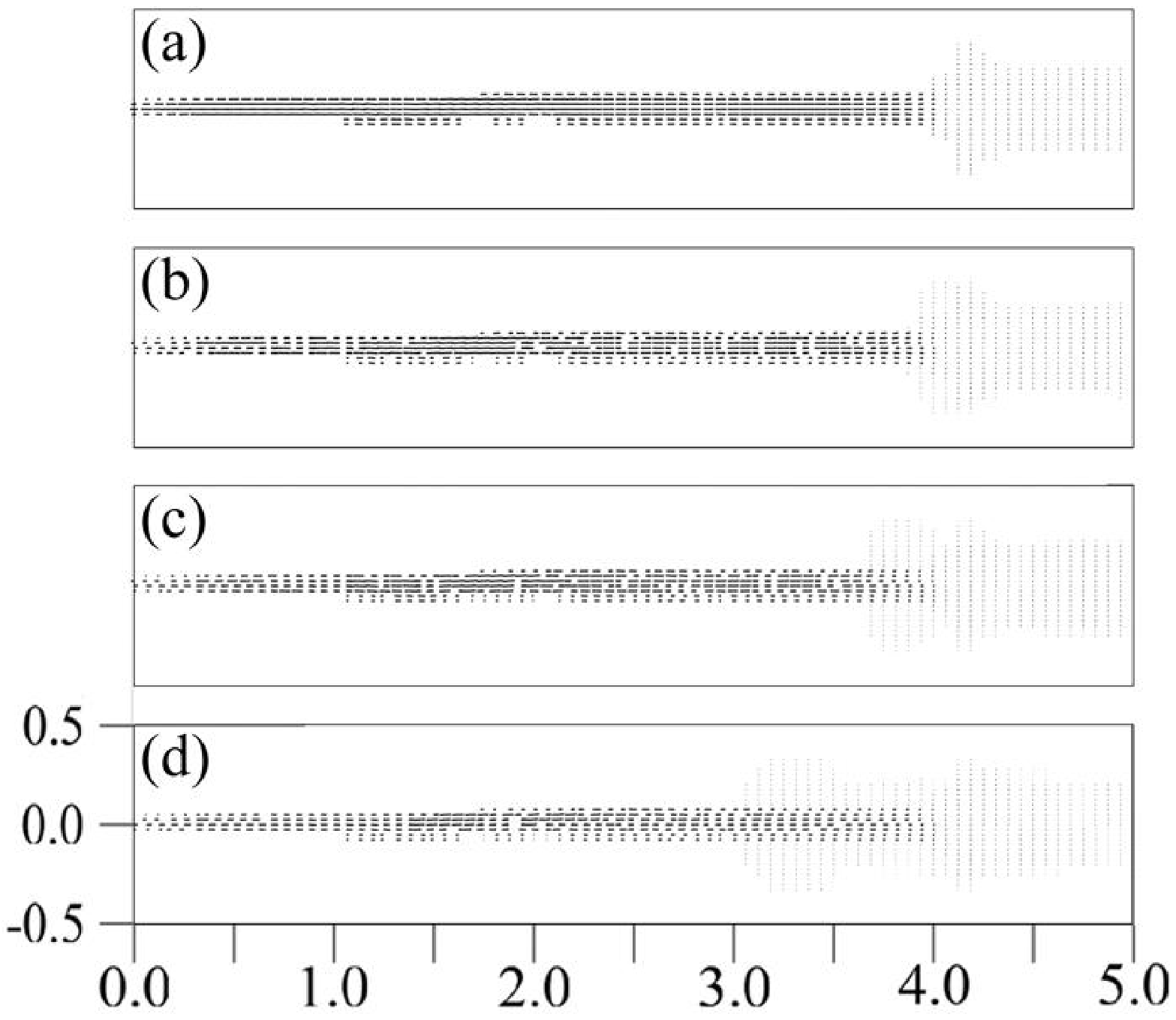}
\caption{Calculated model counterparts for the projected magnetic field,
 when seen at (a) $90^\circ$, (b) $70^\circ$, (c) $55^\circ$, and (d)
 $30^\circ$, from the axis at $t = 15.0$.
 The bars represent the direction of the projected magnetic field and
 the length of the bars represents the polarization intensity.
\label{FIG11}}
\end{figure}

Figure \ref{FIG10} shows the
total intensity distributions. When the angle between the jet axis and
the line-of-sight, $\theta$, is equal to $90^\circ$, total intensity has
the highest value on the jet axis since both the emissivity and the
integration depth of the emitter are maximum. As the viewing
angle becomes small, the asymmetry appears owing to the asymmetry of
the emissivity, this is equivalent to the asymmetry of the magnetic
field component perpendicular to the line-of-sight. The jet has the
helical magnetic field, therefore it is opposite in both sides of the
jet whether the emissivity increases or decreases when the viewing angle
changes. Figure \ref{FIG11} shows the
distributions of the projected magnetic field. It is parallel to the jet
axis in almost all region independent of the viewing angle because the
pitch angle of the helical magnetic field is not so large. However, the
polarization intensity becomes lower as $\theta$ becomes small.

\subsection{After the formation of the wiggled structure}
\begin{figure}
\figurenum{12}
\epsscale{0.77}
\plotone{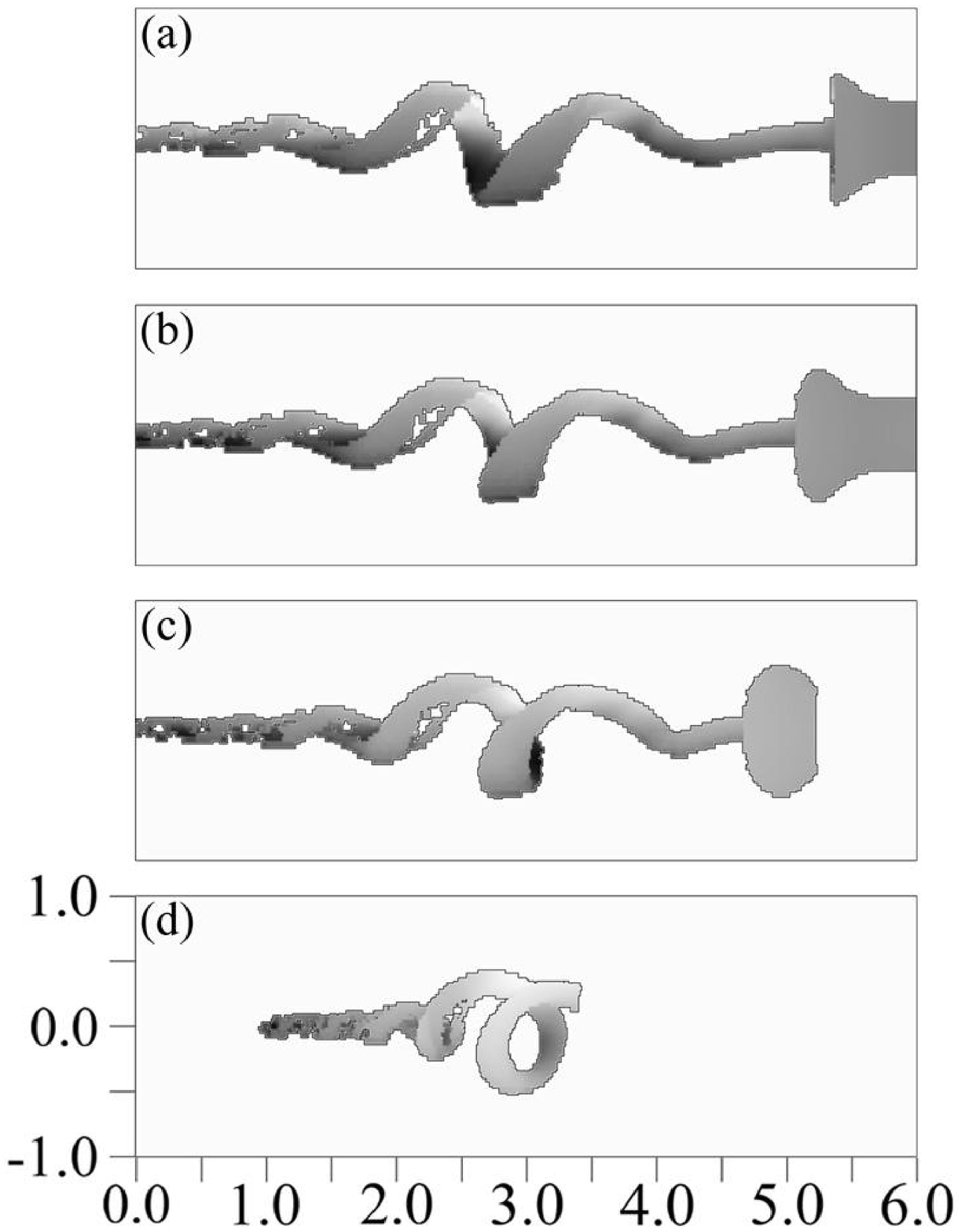}
\caption{Calculated model counterparts for the Faraday rotation measure
 distribution, when seen at (a) $90^\circ$, (b) $70^\circ$, (c)
 $55^\circ$, and (d) $30^\circ$, from the axis (ahead of the jet) at the
 time after the formation of the wiggled structure ($t = 20.0$).
 The data of the numerical observation exists inside the area surrounded
 by the black line.
\label{FIG12}}
\end{figure}

The FRM distributions at the time after the formation of the wiggled
structure ($t = 20.0$) are shown in Figure \ref{FIG12}. The gradient
across the jet axis is partly seen, but not seen in the wiggled structure
because the toroidal component of the magnetic field is smaller compared
to the field component along the structural axis; i.e., the magnetic
field is less twisted around the wiggled structure. Therefore FRM is
large in
the structure coming toward us and small in the structure going
backward. This means that the FRM distribution in the wiggled structure
can strongly reflect its 3-dimensional shape. FRM changes suddenly
where different parts of the wiggled jet overlap each other as seen by
the observer.

\begin{figure}
\figurenum{13}
\epsscale{0.77}
\plotone{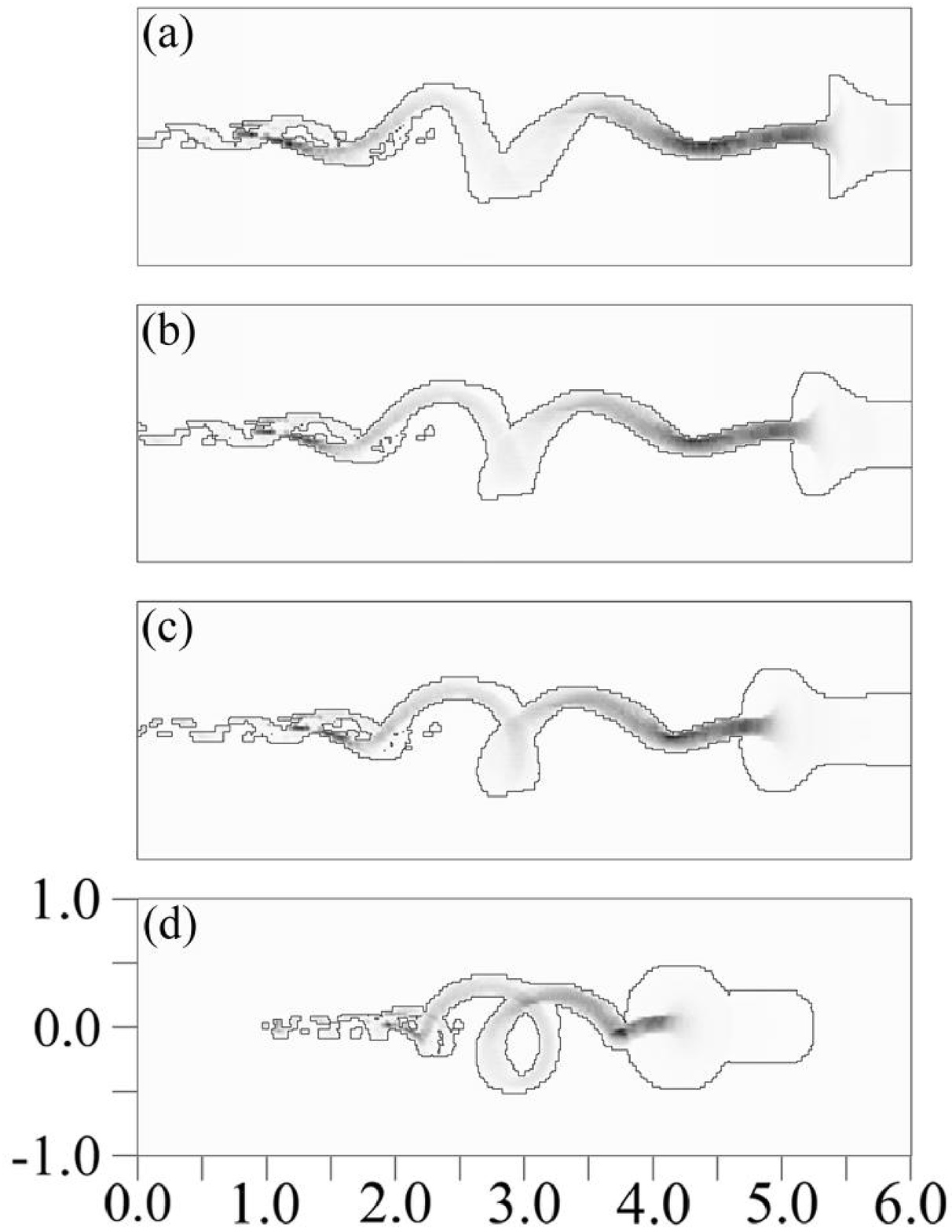}
\caption{Calculated model counterparts for the total intensity, when
 seen at (a) $90^\circ$, (b) $70^\circ$, (c) $55^\circ$, and (d)
 $30^\circ$, from the axis at $t = 20.0$.
 The data of the numerical observation exists inside the area surrounded
 by the black line.
\label{FIG13}}
\end{figure}

\begin{figure}
\figurenum{14}
\epsscale{0.77}
\plotone{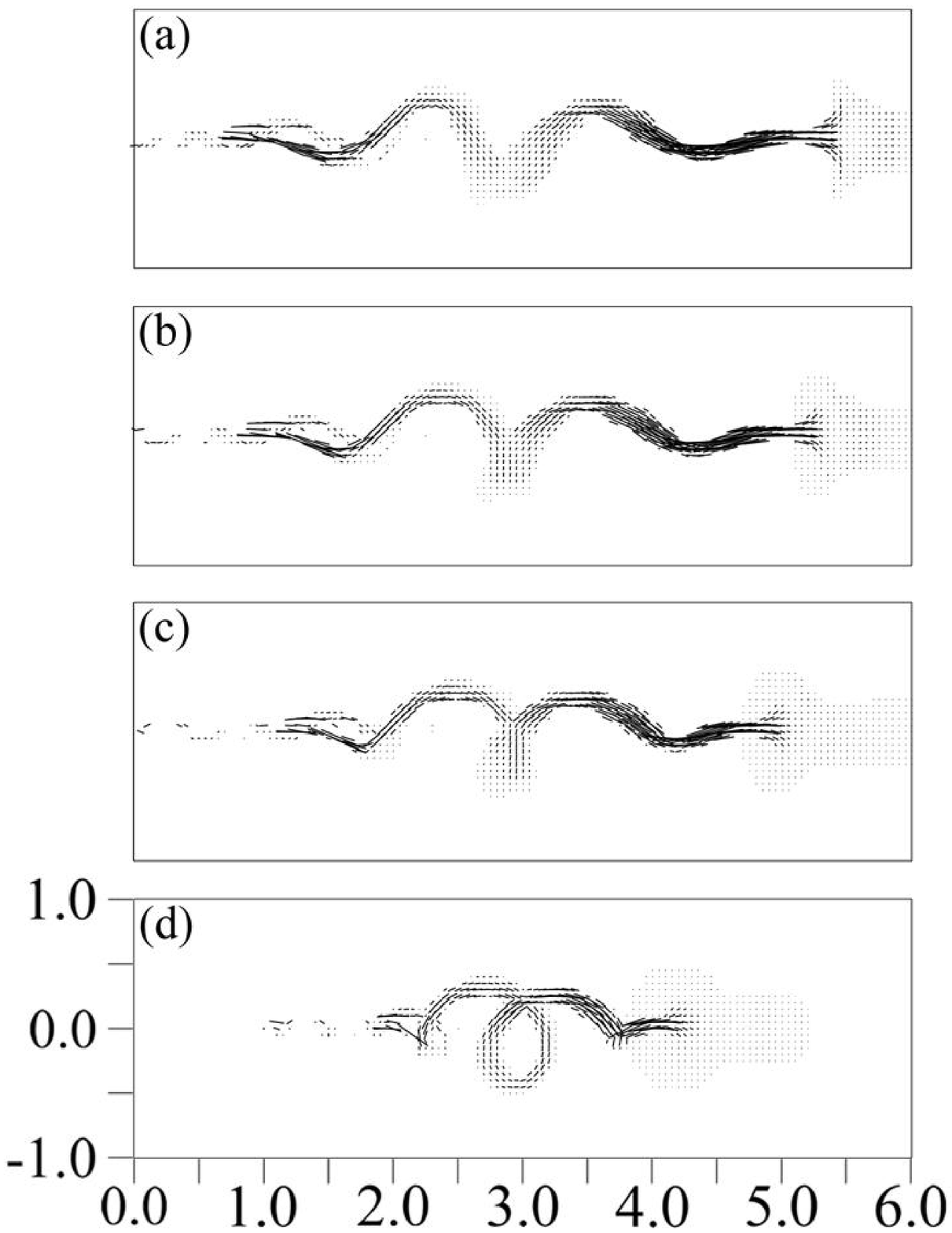}
\caption{Calculated model counterparts for the projected magnetic field,
 when seen at (a) $90^\circ$, (b) $70^\circ$, (c) $55^\circ$, and (d)
 $30^\circ$, from the axis at $t = 20.0$.
 The bars represent the direction of the projected magnetic field and
 the length of the bars represents the polarization intensity.
\label{FIG14}}
\end{figure}

Figure \ref{FIG13} shows the total intensity
distributions. Behind the fast-mode MHD shock front, toroidal magnetic
field is accumulated and magnetic field is strengthened. It leads the
emissivity to increase and the total intensity to become high. On the
other hand, the intensity is not so high in the wiggled structure, where
the magnetic field is less twisted around the wiggled
structure. However, as the viewing angle becomes
small, the intensity in the wiggled structure increases because the
structure which lies on this side and that on the back side overlap
on the same ray. This means that the integration depth of the emitter
increases. For the same reason, the polarization intensity gets higher
in the wiggled structure (Figure \ref{FIG14}). The projected magnetic
field is along the structural axis of the jet on the whole. 
In the wiggled structure, the projected field appears transverse to the 
global jet axis ($z$-axis).

\section{Discussion}
\begin{figure}
\figurenum{15}
\epsscale{1.0}
\plotone{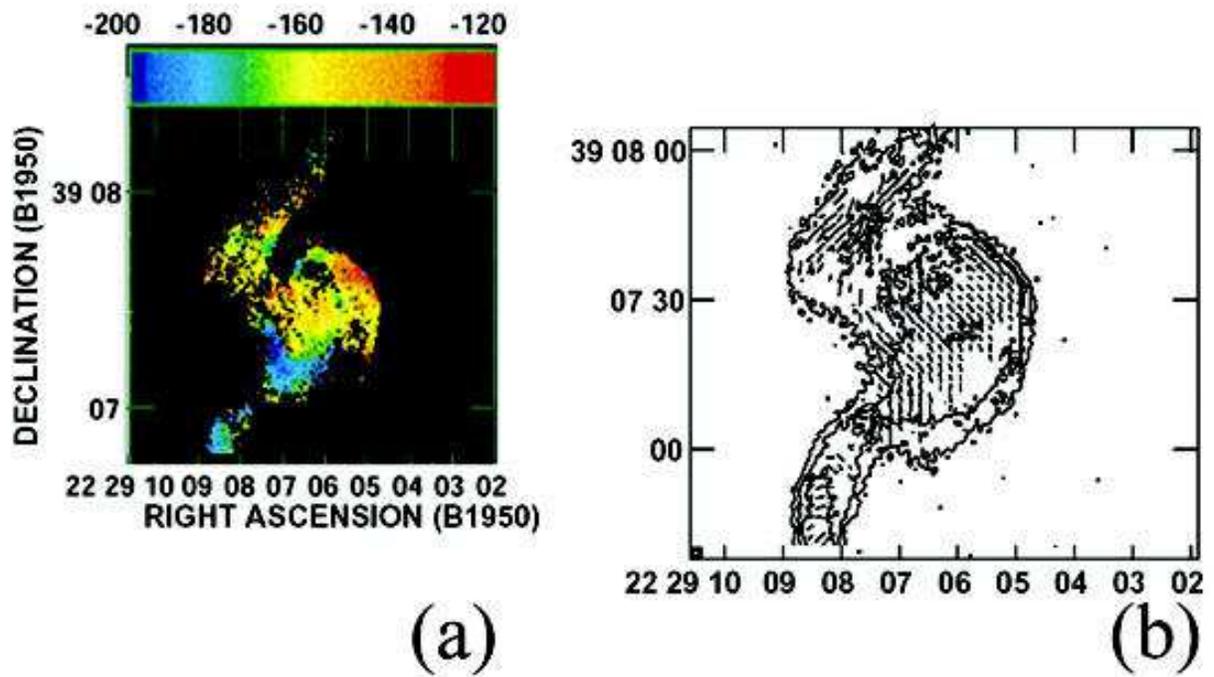}
\caption{The distributions of (a) Faraday rotation measure and (b)
 the projected magnetic field in part of the 3C449 jet (Feretti et
 al. 1999).
\label{FIG15}}
\end{figure}

In this section, we discuss the comparison between the calculated
distributions shown in the previous section and those in the 3C449 jet. Figure
\ref{FIG15} shows the distributions of FRM and the projected
magnetic field in part of the 3C449 jet (Feretti et al. 1999). As we go
along the jet, the color in the FRM distribution starts from blue,
changes to deep blue, jumps to yellow, and changes to red at an edge of
the jet (a bending point) on the right hand side. The parallel
projected magnetic field exists widely where the color in the FRM
distribution is yellow or red.

\begin{figure}
\figurenum{16}
\epsscale{1.0}
\plotone{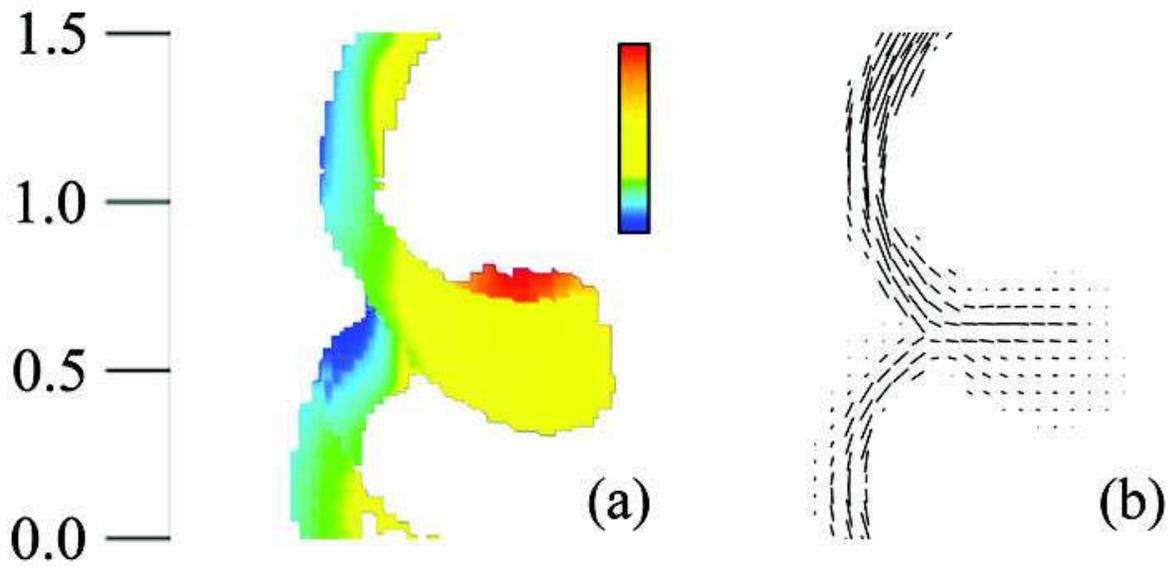}
\caption{The distributions of (a) Faraday rotation measure and (b)
 the projected magnetic field calculated from the results of the
 numerical simulation at $t = 20.0$, seen at $55^\circ$ from the jet
 axis. In terms of the comparison with the observation of the 3C449 jet,
 the length unit of the simulation corresponds to $1'$.
\label{FIG16}}
\end{figure}

\begin{figure}
\figurenum{17}
\epsscale{0.5}
\plotone{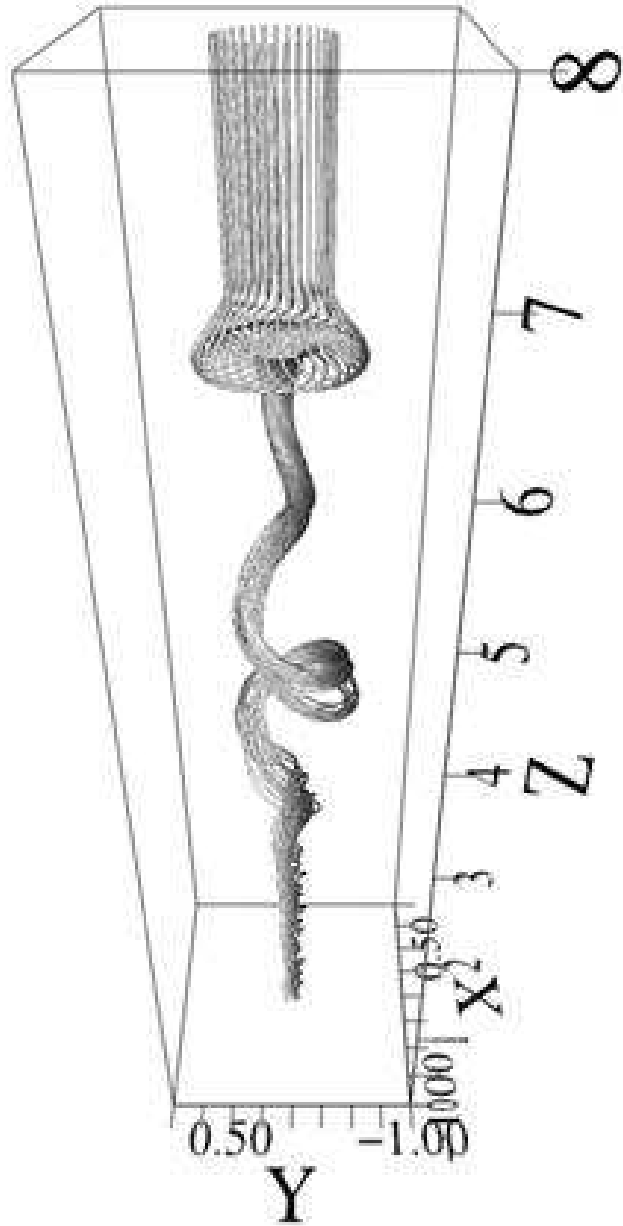}
\caption{The 3-dimensional configuration of the selected magnetic
 lines of force seen at the best-fitting viewing angle at $t = 20.0$.
\label{FIG17}}
\end{figure}

Figure \ref{FIG16} shows the color FRM distribution of the Figure
\ref{FIG11}(c) and the projected magnetic field distribution ($\theta$
is equal to $55^\circ$) in the wiggled structure. The structure takes
the form of a 3-dimensional helix. The structure going backward
corresponds to blue, while that coming toward us corresponds to red. The
structure going upward almost perpendicularly to the line-of-sight
corresponds to the yellow part. The projected magnetic field in the
wiggled structure is almost parallel. This is caused by the following;
the structure going from left to right on the back side and that from
right to left on this side appear to overlap partly (cf. Figure
\ref{FIG17}).

\begin{figure}
\figurenum{18}
\epsscale{0.5}
\plotone{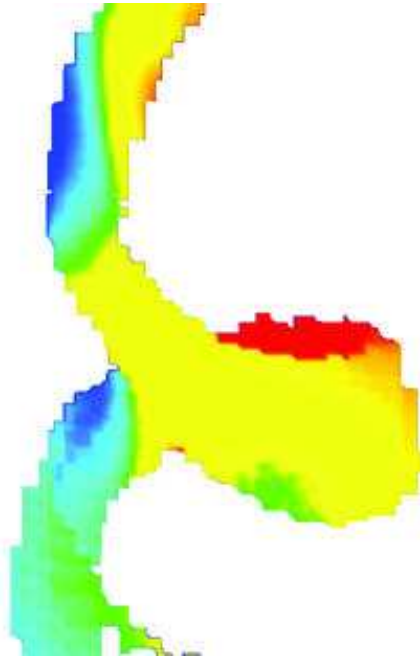}
\caption{The distribution of Faraday rotation measure calculated by the
 integration of $n_e B_\parallel$.
\label{FIG18}}
\end{figure}

As stated in section 3, we omitted the effect of the dense shell from
consideration. Here we demonstrate the effect of the dense shell. Figure
\ref{FIG18} shows the FRM distribution in the wiggled structure calculated
by the integration of $n_e B_\parallel$. The characteristics of the
distribution are same so that it can be said that the dense shell do not
affect the distribution qualitatively so much in this case.

Feretti et al. (1999) estimated the jet velocity and the inclination of
the line-of-sight by the ratio of $P_{c-obs}$ to $P_{c-exp}$, where
$P_{c-obs}$ is the really observed core radio power at 5GHz and
$P_{c-exp}$ is the power inferred from the total radio power at 408MHz.
From such estimation, they concluded that an initial jet velocity
is $0.9c$ and a jet inclination to the line-of-sight is
$82.5^\circ$. However, these values were obtained from the information
about near the core. It is likely that the jet is bent in some
places. Birkinshaw, Laing, \& Peacock (1981) estimated from the
polarization intensity along the jet that the inclination of the north
jet in 3C449 is $53^\circ$. This inclination is almost same with that at
which we can reproduce the characteristics of the observation.

These similarities between the calculated distributions and the observed
ones suggest that the 2-dimensional wiggle in
the observation is due to the projection of a 3-dimensional helical
structure onto the plane of the sky. It is also suggested that
the structure which lies on this side (yellow) hides the structure
on the back side. The reddish part corresponds to the part coming toward
us in the helix. If this is the case, it is highly likely that the
magnetic field (especially the toroidal component) plays an important
role not only in the emission but also in the dynamics of the jet
formation because the wiggle formation is caused by the accumulation of
the toroidal field. It is also suggested that the FRM distribution can
be a sensitive clue for telling which of the proposed models is the
correct one.

Because it is not obvious that how much the Faraday screen that is 
irrelevant to the jet itself contributes to the value of rotation measure, 
it is not possible to predict the value of Faraday rotation measure. It is, 
however, possible to estimate the difference between the maximum and the 
minimum of rotation measure. In our simulation, the scale of the wiggled 
structure is almost equal to the length unit of the simulation. 
Integration depth of Faraday screen is about twice as long as the 
length unit. The rotation measure is given by FRM [rad m$^{-2}$] = $8.1 
\times 10^5 n_e$ [cm$^{-3}$] $B_\parallel$ [G] $L_{FS}$ [pc], where 
$n_e$ is the electron density, $B_\parallel$ is the line-of-sight component 
of the magnetic field and $L_{FS}$ is the integration depth of the Faraday 
screen. In our calculation, $L_{FS}$ is twice as long as $L_{wiggle}$ where 
$L_{wiggle}$ is the scale of the wiggled structure. If the maximum of the 
rotation measure (FRM$_{max}$) is made by $B_\parallel = \langle B \rangle$ 
and the minimum of the rotation measure (FRM$_{min}$) by $B_\parallel = - 
\langle B \rangle$, then FRM$_{max}$ - FRM$_{min}$ $= 3.2 \times 10^6 n_e 
\langle B \rangle L_{wiggle}$. In the case of 3C449 jet, the scale of the 
wiggled structure is about $1'$ (from $39^\circ07'$ to $39^\circ08'$ in Figure 
\ref{FIG15}) and it corresponds to 30kpc. Finally FRM$_{max}$ - FRM$_{min}$ 
is equal to $9.8 \times 10^{10} n_e \langle B \rangle$. In the case of 3C449 
jet, FRM$_{max}$ - FRM$_{min}$ is nearly equal to 100 [rad m$^{-2}$] (Feretti 
et al. 1999). It is therefore predicted from our calculation that $n_e 
\langle B \rangle$ in the jet is equal to $10^{-9}$ [G cm$^{-3}$].

Except for the wiggled structure, there are some differences. First, the 
total and polarized intensities enhance beyond the wiggled structure in the 
model counterparts while it is not seen in the observation. The enhancement 
of the intensities in the model counterparts is due to the accumulation of 
the toroidal magnetic field. This is equivalent to the enhancement of the 
current density. It is therefore likely that the magnetic reconnection takes 
place. The magnetic reconnection is the dissipation of the magnetic field so 
that the strength of the magnetic field decreases and the intensities can 
also decrease. For more exact discussion, resistive MHD simulations are 
necessary. Second, the projected magnetic field is longitudinal at the side 
near the core in the model counterparts while it is transverse in the 
observation. It is remarkable that the transverse field was obtained in 
paper I in which the part of the jet near the core was dealt with. We 
therefore think that the simulations simultaneously solving the formation, 
propagation and destabilization of the jet are necessary.

It can be said that the influence of the establishment of
the validity of the magnetic model, especially the ``sweeping magnetic 
twist model'', will be extremely large. This claims that the dynamics of
the jet formation, including the collimation and the destabilization in
large scale, is due to the operation of the magnetic field. Since there
is no local source for this large energy to do the job in the part of
the space where the wiggles of the jets are seen (closer to the hotspots
at the tip of long jets), the energy should come all the way from the
AGN core. We also claim that the energy dumped in the hotspots and
the radio lobes is carried by the Poynting flux. 
In MHD models, the toroidal component is generated and its propagation
carries the Poynting energy, therefore the models are plausible. The
generation of the toroidal magnetic field may be due to either the
rotation of the accretion disk (Blandford \& Payne 1982, Uchida \&
Shibata 1985) or the effect of the rotating black hole
(Blandford \& Znajek 1977, Koide et al. 2000). This means that the large
amount of energy is 
flowing out from the core part of an AGN (though we can not yet say
exactly whether that comes from the inner edge of the accretion disk, or
from the black hole). If we accept this, the major bearer of energy
flowing out is, even in the AGN core itself, large amplitude TAWs, and
even very high Lorentz factor phenomena will be a product of the
energetic TAWs.

\section{Conclusion}
We performed a 3-dimensional MHD simulation based on our ``sweeping
magnetic twist model'', which was applied to the situation far from the
AGN core by Nakamura et al. (2001). In the ``sweeping magnetic twist
model'', the disk rotation generates the toroidal magnetic field and it
propagates into two directions along a large scale magnetic field as
torsional Alfv\'en waves (TAWs). In this paper, the situation TAWs are
propagating far from the gravitator was dealt with. It was assumed that
there is a lower Alfv\'en velocity region ahead of the propagating TAWs.
The toroidal magnetic field becomes accumulated after TAWs enter the low
Alfv\'en velocity region. This causes the growth of the current-driven
helical kink instability and the wiggled structure is formed.

We also calculated the observational quantities (FRM, Stokes, $I$, $Q$,
and $U$ parameters) by integrating the numerical data along the
line-of-sight. The Faraday rotation screen and the emitting
region were defined separately. The integration for FRM was done only in
the Faraday rotation screen and that for Stokes parameters only in the
emitting region. The projected magnetic field was determined from $Q$
and $U$.

Before the formation of the wiggled structure, the FRM distribution has
a gradient across the jet axis. An asymmetry in the total intensity
distribution exists except the case where $\theta$ is equal to
$90^\circ$. Same features are seen in the result of paper I and these
features can explain some observations (Perley et al. 1984, Asada et
al. 2002, Gabuzda \& Murray 2003). The projected magnetic field is
parallel to the jet axis because the pitch angle of the helical magnetic
field is not so large.

After the formation of the wiggled structure, the FRM distribution has
a clear correlation with the large scale structure of the jet
itself. This is caused by the following; the magnetic field becomes less
twisted around the wiggled structure owing to the current-driven helical
kink instability. This is equivalent to that the magnetic field is
almost along the structural axis of the jet in the wiggled
structure. The total and polarization intensities are
low in the wiggled structure when the jet inclination to the
line-of-sight is large. However, they become higher as the jet
inclination becomes small because the structure on the backside and that
on the frontside of the 3-dimensional helix overlap on the same ray as
seen by observer. We found that when we see the jet at a certain angle,
we can reproduce the characteristics of the observation of the 3C449 jet
(Feretti et al. 1999).
This suggests that the FRM distribution could be strongly affected by
the magnetic field in or around the jet and that the jet has a helical
structure. If this is the case, it is also suggested that the magnetic
field (especially the toroidal component) plays an important role in the
formation of astrophysical jets.

\acknowledgments
Finally, we would like to thank Dr. L. Feretti for providing their
precious data of FRM, and for valuable discussions. One of the
authors (H. K.) thanks Syuniti Tanuma and Troels Haugb{\o}lle for helpful
comments. We would hope that more 
observations of the FRM distribution in AGN jets will be performed in
the near future because they are very important and scientifically
rewarding. This means that with those we can determine the correct model
if our claim in the above is correct. We would therefore urge high
quality observations of the FRM distribution, to seek for the clues of
the mechanism of jet formation. Numerical computations were carried out
on VPP5000 at the Astronomical Data Analysis Center of the National
Astronomical Observatory, Japan, which is an interuniversity research
institute of astronomy operated by the Ministry of Education, Culture,
Sports, Science, and Technology.

 

\clearpage

\clearpage

\clearpage

\clearpage

\clearpage

\clearpage

\clearpage

\clearpage

\clearpage

\clearpage

\clearpage

\clearpage

\clearpage

\clearpage

\clearpage

\clearpage

\clearpage

\clearpage

\end{document}